\documentclass[12pt]{article}
\usepackage{amsmath,amssymb}
\usepackage{bm}
\usepackage{mathrsfs}
\usepackage{fourier}
\usepackage{multirow}
\allowdisplaybreaks[4]
\newcommand{{\SlashD}}{D\!\!\!\!\!\!\big/}
\newcommand{{\Slashq}}{q\!\!\!\!\!\big/}
\newcommand{{\SlashF}}{{\rm F}\!\!\!\!\!/}
\usepackage[usenames]{color}

\begin{document}

\title{On a mechanism realizing quark mass hierarchy}

\author{
Yoshiharu \textsc{Kawamura}\footnote{E-mail: haru@azusa.shinshu-u.ac.jp}\\
{\it Department of Physics, Shinshu University, }\\
{\it Matsumoto 390-8621, Japan}\\
}


\maketitle
\begin{abstract}
We reconsider a generation of up-type quark mass hierarchy 
in the standard model
and clarify how a mechanism works to realize the hierarchy
without severe fine tuning.
\end{abstract}

\section{Introduction}

It is expected that the fermion mass hierarchies in the standard model (SM) 
are elegantly understood from unknown features,
e.g., flavor symmetries and a structure behind texture zeros,
based on the top-down approach~\cite{CEG,F,HHW,F2,GJ,FN}.
The flavor structure of quarks and leptons has been studied intensively, 
using various flavor symmetries~\cite{FN,MY,I,HPS,HS,KY,AF,IKOSOT,IKOOST}.
It is important not only to identify flavor symmetries 
but also to figure out what's behind their breaking,
because no exact flavor-dependent symmetries exist in the SM~\cite{LNS,Koide}.
The bottom-up approach has also been used~\cite{Ku,AN,St,DR,YK,YK2,YK3}.

It is mostly believed that the fermion mass hierarchies are derived 
without severe fine tuning.
In other words, tiny Yukawa couplings should not appear
as a result of fine tuning among parameters of $O(1)$ size.
Hence, it is conjectured that 
there exist tiny parameters at a more fundamental level,
and an excellent mechanism works 
to generate a hierarchical structure on physical parameters.

In this letter, we reconsider a generation of 
up-type quark mass hierarchy in the SM
and clarify how a mechanism works to realize the hierarchy
without severe fine tuning, based on the following ideas.
The SM is an effective quantum field theory (QFT),
and QFT is a tool or a framework to describe quantum phenomena 
in an efficient manner.
Parameters in the SM Lagrangian are not necessarily fundamental 
but effective ones, and a tininess of some physical parameters
is naturally understood by using more fundamental parameters.

The outline of this letter is as follows.
In the next section, we present our basic idea in general terms.
We examine up-type quark mass hierarchy in Sect. 3.
In the last section, we give conclusions and discussions.

\section{Basic idea}

We show that a hierarchy among physical parameters in magnitude 
can be generated at a tree level without severe fine tuning,
when more fundamental parameters exist.
We study an effective QFT described by the Lagrangian density:
\begin{eqnarray}
\mathscr{L}_{\rm QFT} 
= \sum_{\alpha} \frac{c_{\alpha}}{\varLambda^{d_{\alpha}-4}} O_{\alpha},
\label{L-QFT}
\end{eqnarray}
where $c_{\alpha}$ are dimensionless parameters,
$\varLambda$ is a cutoff scale,
and $O_{\alpha}$ are operators with mass dimensions $d_{\alpha}$.
We assume that particles are
weakly coupled, i.e., $|c_{\alpha}| \le O(1)$.

Let $c_{\alpha}$ be classified into two categories.
One is a set of parameters that are physical in itself, 
e.g., gauge couplings in the SM.
The other is a set of parameters where physical parameters 
$\tilde{C}_a$ originate from, 
after redundant ones (unphysical ones) are eliminated, 
e.g., Yukawa coupling matrices $y_{ij}^{(f)}$ in the SM. 
Hereafter, we focus on a second one.
When there appears a hierarchy among $\tilde{C}_a$ in magnitude
or $\tilde{C}_a$ contain a tiny parameter such as up quark Yukawa coupling $y_u$,
a cancellation among $c_{\alpha}$ is, in general, needed
to derive a tiny one with $|\tilde{C}_a| \ll O(1)$ from $|c_{\alpha}| = O(1)$.
In the following, we analyze this feature in a quantitative way.

First, we regard $c_{\alpha}$ as fundamental parameters,
and introduce a measure of fine tuning defined by\footnote{
This type of measure is originally proposed to quantify the degree of 
fine tuning on the Higgs boson mass
among soft supersymmetry breaking parameters~\cite{BG}.
}
\begin{eqnarray}
\varDelta_{c_{\alpha}}\tilde{C}_a \equiv 
\left|\frac{\partial \ln \tilde{C}_a}{\partial \ln c_{\alpha}}\right|
= \left|\frac{\partial \tilde{C}_a}{\partial {c}_{\alpha}}\frac{c_{\alpha}}{\tilde{C}_a}\right|.
\label{D}
\end{eqnarray}
Here and hereafter no summations on $\alpha$ and $a$ are done.
The value of $\varDelta_{c_{\alpha}}\tilde{C}_a$
implies a necessary cancellation of first part of that.
Or the smaller $\varDelta_{c_{\alpha}}\tilde{C}_a$ are, the less a degree of fine tuning is.
For instance, in a case with 
$\displaystyle{\left|{\partial \tilde{C}_a}/{\partial {c}_{\alpha}}\right| = O(1)}$
and $|c_{\alpha}| \gg |\tilde{C}_a|$ for some $\tilde{C}_a$,
we have $\varDelta_{c_{\alpha}}\tilde{C}_a \gg O(1)$
and need severe fine tuning among $c_{\alpha}$ to obtain $\tilde{C}_a$.\footnote{
The dependency of $\tilde{C}_a$ in $c_{\alpha}$
can be restricted by imposing on 
a condition such as $\varDelta_{c_{\alpha}}\tilde{C}_a \le O(1)$.
The finite version is used in a bottom-up approach 
based on `stability' principle~\cite{St,DR}.
Here, the stability principle means that
a tiny parameter should not be sensitive to a change of
fundamental parameters.
}

We give a simple example that $\tilde{C}_a$ are given as linear combinations of
$c_{\alpha}$ such that $\tilde{C}_1 = \sum_{\alpha = 1}^3 A_{1\alpha} c_{\alpha}$
and $\tilde{C}_2 = \sum_{\alpha = 1}^3 A_{2\alpha} c_{\alpha}$.
We assume that $A_{1\alpha}$ and $A_{2\alpha}$ are irrelevant to $c_{\alpha}$,
and their magnitudes are given by
$|A_{1\alpha}|=O(1)$, $|A_{2\alpha}|=O(1)$, and $|c_{\alpha}| = O(1)$.
In the presence of $|\tilde{C}_1| = O(1) \gg |\tilde{C}_2|$,
degrees of fine tuning are estimated as
\begin{eqnarray}
\varDelta_{c_{\alpha}}\tilde{C}_1 
= \left|A_{1\alpha} \frac{c_{\alpha}}{\tilde{C}_1}\right| = O(1),~~
\varDelta_{c_{\alpha}}\tilde{C}_2 
= \left|A_{2\alpha} \frac{c_{\alpha}}{\tilde{C}_2}\right| \gg O(1).
\label{D-Ex}
\end{eqnarray}
The second relation means that severe fine tuning is needed to derive $\tilde{C}_2$.

Next, we consider a case with more fundamental parameters.
They are classified into two categories, $\{f_k\}$
and $\{\varepsilon_l\}$,
based on their magnitudes, i.e., $f_k = O(1)$ and $\varepsilon_l \ll O(1)$.
Here, their values are taken as positive ones.
For example, $f_k$ emerge as vacuum expectation values (VEVs) of 
some scalar fields such as moduli fields $\phi_k^{({\rm M})}$, 
i.e., $f_k = |\langle \phi_k^{({\rm M})} \rangle|/M$.
Here, $M$ is a fundamental scale such as a string scale or the Planck scale.
Then, the magnitudes of $f_k$ are $O(1)$, 
if those of $\langle \phi_k^{({\rm M})} \rangle$ are $O(M)$.
$\varepsilon_l$ are defined by
$\varepsilon_l \equiv |\langle \varphi_l \rangle|/M$,
and their magnitudes are much less than $O(1)$, 
when $|\langle \varphi_l \rangle| \ll M$.

For simplicity, we examine a case with two fundamental parameters $f$ and $\varepsilon$
such that $f = O(1) \gg \varepsilon$.
In this case, $\tilde{C}_a$ are functions of $f$ and $\varepsilon$, 
i.e., $\tilde{C}_a = \tilde{C}_a(f, \varepsilon)$,
and degrees of fine tuning are measured by
\begin{eqnarray}
\varDelta_{f}\tilde{C}_a \equiv 
\left|\frac{\partial \ln \tilde{C}_a}{\partial \ln f}\right|
= \left|\frac{\partial \tilde{C}_a}{\partial f}\frac{f}{\tilde{C}_a}\right|,~~
\varDelta_{\varepsilon}\tilde{C}_a \equiv 
\left|\frac{\partial \ln \tilde{C}_a}{\partial \ln \varepsilon}\right|
= \left|\frac{\partial \tilde{C}_a}{\partial \varepsilon}\frac{\varepsilon}{\tilde{C}_a}\right|.
\label{Dfe}
\end{eqnarray}
If the following conditions fulfill,
\begin{eqnarray}
\left|\frac{\partial \tilde{C}_a}{\partial f}\right| 
\le O\left(\left|{\tilde{C}_a}\right|\right),~~
\left|\frac{\partial \tilde{C}_a}{\partial \varepsilon}\right| 
\le O\left(\frac{\left|\tilde{C}_a\right|}{\varepsilon}\right),
\label{cf-cond}
\end{eqnarray}
$\varDelta_{f}\tilde{C}_a \le O(1)$ and $\varDelta_{\varepsilon}\tilde{C}_a \le O(1)$
are derived, and then tiny parameters can be obtained 
without severe fine tuning.
This feature is understood from a viewpoint of perturbation, i.e.,
effects of $\varepsilon$ can be perturbatively incorporated, as follows.
For $\left|\tilde{C}_a\right| = O(1)$, $\tilde{C}_a$ are, in general, expanded as power series 
of $\varepsilon$:
\begin{eqnarray}
\tilde{C}_a = \sum_{n = 0}^{\infty} \tilde{C}_{a, n}(f) \varepsilon^n,
\label{caO(1)}
\end{eqnarray}
where $\left|\tilde{C}_{a, 0}\right| = O(1)$.
If $\left|\partial \tilde{C}_{a, 0}/\partial f\right| \le O(1)$
and $\left|\tilde{C}_{a, 1}\right| \le O(1)$ hold,
$\varDelta_{f}\tilde{C}_a \le O(1)$ and $\varDelta_{\varepsilon}\tilde{C}_a \le O(1)$
are derived.
For $\left|\tilde{C}_a\right| = O(\varepsilon) \ll O(1)$, 
$\tilde{C}_a$ are expanded as power series 
of $\varepsilon$:
\begin{eqnarray}
\tilde{C}_a = \sum_{n = 1}^{\infty} \tilde{C}_{a, n}(f) \varepsilon^n,
\label{caO(e)}
\end{eqnarray}
where $\left|\tilde{C}_{a, 1}\right| = O(1)$.
In this case, $\left|\partial \tilde{C}_{a}/\partial \varepsilon\right| = O(1)$ holds.
If $\left|\partial \tilde{C}_{a, 1}/\partial f\right| \le O(1)$ holds,
$\varDelta_{f}\tilde{C}_a \le O(1)$ and $\varDelta_{\varepsilon}\tilde{C}_a = O(1)$
are derived.
Note that the term with $n=0$ is missing in Eq.~(\ref{caO(e)}),
and its absence could be due to the existence of some symmetries.

Let us explain the above feature by way of $c_{\alpha} = c_{\alpha}(f, \varepsilon)$.
We assume that $c_{\alpha}(f, \varepsilon)$ originate from 
independent terms in a more fundamental theory, 
and $c_{\alpha}$ are written by
\begin{eqnarray}
c_{\alpha} = c^{(1)}_{\alpha}(f) + c^{(\varepsilon)}_{\alpha}(f, \varepsilon),
\label{c}
\end{eqnarray}
where $|c^{(1)}_{\alpha}| = O(1)$, $|c^{(\varepsilon)}_{\alpha}| = O(\varepsilon)$,
$|\partial c^{(1)}_{\alpha}/\partial f| = O(1)$,
$|\partial c^{(\varepsilon)}_{\alpha}/\partial f| = O(\varepsilon)$
and $|\partial c^{(\varepsilon)}_{\alpha}/\partial \varepsilon| = O(1)$.
Then, $\varDelta_{f}\tilde{C}_a$ and $\varDelta_{\varepsilon}\tilde{C}_a$
are calculated as 
\begin{eqnarray}
\varDelta_{f}\tilde{C}_a 
= \left|\sum_{\alpha} \left(\frac{\partial \tilde{C}_a}{\partial c_{\alpha}^{(1)}}
\frac{\partial c_{\alpha}^{(1)}}{\partial f} \frac{f}{\tilde{C}_a}
+ \frac{\partial \tilde{C}_a}{\partial c_{\alpha}^{(\varepsilon)}}
\frac{\partial c_{\alpha}^{(\varepsilon)}}{\partial f} \frac{f}{\tilde{C}_a}\right)\right|,~~
\varDelta_{\varepsilon}\tilde{C}_a 
= \left|\sum_{\alpha} \frac{\partial \tilde{C}_a}{\partial c_{\alpha}^{(\varepsilon)}}
\frac{\partial c_{\alpha}^{(\varepsilon)}}{\partial \varepsilon} 
\frac{\varepsilon}{\tilde{C}_a}\right|,
\label{Dcfe}
\end{eqnarray}
respectively.
If the following conditions fulfill,
\begin{eqnarray}
\left|\sum_{\alpha} \frac{\partial \tilde{C}_a}{\partial c_{\alpha}^{(1)}}
\frac{\partial c_{\alpha}^{(1)}}{\partial f}\right| 
\le O(|\tilde{C}_a|),~~
\left|\frac{\partial \tilde{C}_a}{\partial c_{\alpha}^{(\varepsilon)}}\right| 
\le O\left(\frac{|\tilde{C}_a|}{\varepsilon}\right),
\label{c-cond}
\end{eqnarray}
we find that $\varDelta_{f}\tilde{C}_a \le O(1)$
and $\varDelta_{\varepsilon}\tilde{C}_a \le O(1)$.
The point is that $\tilde{C}_a$ are not functions of $c_{\alpha}$
but $c_{\alpha}^{(1)}$ and $c_{\alpha}^{(\varepsilon)}$, or
$c_{\alpha}^{(1)}$ and $c_{\alpha}^{(\varepsilon)}$ are treated as independent ones.

We reconsider the previous example that
$\tilde{C}_1 = \sum_{\alpha = 1}^3 A_{1\alpha} c_{\alpha}$
and $\tilde{C}_2 = \sum_{\alpha = 1}^3 A_{2\alpha} c_{\alpha}$
with $|A_{1\alpha}|=O(1)$, $|A_{2\alpha}|=O(1)$, and $|c_{\alpha}| = O(1)$.
If $c_{\alpha}$ are given by Eq.~(\ref{c}) 
and $\sum_{\alpha =1}^3 A_{2\alpha}c^{(1)}_{\alpha} = 0$ holds,
we obtain the hierarchy in magnitude such that
\begin{eqnarray}
|\tilde{C}_1| = \left|\sum_{\alpha = 1}^3 \left(A_{1\alpha} c_{\alpha}^{(1)}(f)
+ A_{1\alpha} c_{\alpha}^{(\varepsilon)}(f, \varepsilon)\right)\right| = O(1),~~
|\tilde{C}_2| = \left|\sum_{\alpha = 1}^3 
A_{2\alpha} c_{\alpha}^{(\varepsilon)}(f, \varepsilon)\right| = O(\varepsilon).
\label{tilde{c}}
\end{eqnarray}
The absence of $c_{\alpha}^{(1)}$ in $\tilde{C}_2$
could be related to some symmetries.

In this way, we arrive at the idea that
{\it parameters in an effective QFT are not necessarily fundamental but effective ones,
and then a tininess of some physical parameters
can be naturally understood by using more fundamental parameters.}

Strictly speaking, we decipher 
a mechanism that a hierarchy among physical parameters 
can be realized {\it at a tree level} without severe fine tuning.
Physical parameters, in general, receive radiative corrections,
and hence we need to examine whether the hierarchy is stabilized
against radiative corrections or not.
Some symmetries can play a central role to 
the stabilization, and we suppose that they function effectively in our case.

\section{Consideration of fine tuning on Yukawa couplings}

\subsection{Quark Yukawa couplings}

The quark Yukawa coupling matrices $y^{(u)}$ and $y^{(d)}$ 
are diagonalized bi-unitary transformations as
\begin{eqnarray}
&~& V_{\rm L}^{(u)} y^{(u)} {V_{\rm R}^{(u)}}^{\dagger} 
= y_{\rm diag}^{(u)} = {\rm diag}\left(y_u, y_c, y_t\right),
\label{yu-diag}\\
&~& V_{\rm L}^{(d)} y^{(d)} {V_{\rm R}^{(d)}}^{\dagger}
= y_{\rm diag}^{(d)} = {\rm diag}\left(y_d, y_s, y_b\right),
\label{yd-diag}
\end{eqnarray}
where $V_{\rm L}^{(u)}$, $V_{\rm L}^{(d)}$, $V_{\rm R}^{(u)}$ and $V_{\rm R}^{(d)}$
are unitary matrices, 
and $y_u$, $y_c$, $y_t$, $y_d$, $y_s$ and $y_b$ are Yukawa couplings
of up, charm, top, down, strange, and bottom quarks, respectively.
The Kobayashi--Maskawa (KM) matrix is defined by~\cite{KM}
\begin{eqnarray}
V_{\rm KM} \equiv V_{\rm L}^{(u)} {V_{\rm L}^{(d)}}^{\dagger}.
\label{VKM}
\end{eqnarray}

Each quark mass is obtained by multiplying each Yukawa coupling by
the VEV of neutral component of Higgs doublet.
From Eqs.~(\ref{yu-diag}), (\ref{yd-diag}) 
and experimental values of quark masses,
$y_{\rm diag}^{(u)}$, $y_{\rm diag}^{(d)}$,
and $V_{\rm KM}$ are roughly estimated 
at the weak scale as~\cite{PDG}
\begin{eqnarray}
&~& y_{\rm diag}^{(u)} = 
{\rm diag}\left(1.3 \times 10^{-5},~ 7.3 \times 10^{-3},~ 1.0\right) 
= {\rm diag}\left(\lambda^7, \lambda^4, 1\right),
\label{yu-diag-value}\\
&~& y_{\rm diag}^{(d)} 
= {\rm diag}\left(2.7 \times 10^{-5},~ 
5.5 \times 10^{-4},~ 2.4 \times 10^{-2}\right)
= {\rm diag}\left(\lambda^7, \lambda^5, \lambda^3\right),
\label{yd-diag-value}\\
&~&  V_{\rm KM} = \left(
\begin{array}{ccc}
1 & \lambda & \lambda^4 \\
\lambda & 1 & \lambda^2 \\
\lambda^3 & \lambda^2 & 1
\end{array}
\right),
\label{VKM-lambda}
\end{eqnarray}
where $\lambda^n$ means 
$\displaystyle{O\left(\lambda^n\right)}$
with $\lambda = \sin\theta_{\rm C} \cong 0.225$ 
($\theta_{\rm C}$ is the Cabibbo angle~\cite{C}).

Information on physics beyond the SM is hidden
in $V_{\rm L}^{(u)}$,
$V_{\rm R}^{(u)}$, and $V_{\rm R}^{(d)}$ besides 
observable parameters $y_{\rm diag}^{(u)}$, $y_{\rm diag}^{(d)}$,
and $V_{\rm KM}$.
The matrices $V_{\rm L}^{(u)}$,
$V_{\rm R}^{(u)}$, and $V_{\rm R}^{(d)}$
are completely unknown in the SM,
because they can be eliminated by the global 
${\rm U}(3) \times {\rm U}(3) \times {\rm U}(3)/{\rm U}(1)$ symmetry
that the quark kinetic term possesses.

Using $V_{\rm L}^{(u)}$ and $V_{\rm L}^{(d)}$,
the Hermitian matrices $y^{(u)} {y^{(u)}}^{\dagger}$ and $y^{(d)} {y^{(d)}}^{\dagger}$
are diagonalized by unitary transformations:
\begin{eqnarray}
V_{\rm L}^{(u)} \left(y^{(u)} {y^{(u)}}^{\dagger}\right) {V_{\rm L}^{(u)}}^{\dagger} 
= \left(y_{\rm diag}^{(u)}\right)^2,~~
V_{\rm L}^{(d)} \left(y^{(d)} {y^{(d)}}^{\dagger}\right) {V_{\rm L}^{(d)}}^{\dagger}
= \left(y_{\rm diag}^{(d)}\right)^2.
\label{f-diag^2}
\end{eqnarray}
If Yukawa coupling matrices are specified,
we can obtain $V_{\rm L}^{(u)}$
and $V_{\rm L}^{(d)}$ from (\ref{f-diag^2}) 
and check whether they provide correct KM matrices or not.

As seen from (\ref{yu-diag}), $y_u$, $y_c$, and $y_t$ are written 
as linear combinations of $y^{(u)}$:
\begin{eqnarray}
y_u = \sum_{i, j} {R_{11}}^{ij} y^{(u)}_{ij},~~
y_c = \sum_{i, j} {R_{22}}^{ij} y^{(u)}_{ij},~~
y_t = \sum_{i, j} {R_{33}}^{ij} y^{(u)}_{ij},
\label{yu-linear}
\end{eqnarray}
where $i, j (=1, 2, 3)$ are family labels
and ${R_{i'j'}}^{ij} = \left(V_{\rm L}^{(u)}\right)_{i'i}
\left({V_{\rm R}^{(u)}}^{\dagger}\right)_{jj'}$.
When we regard $y^{(u)}_{ij}$ as fundamental parameters,
a large cancellation seems likely necessary to obtain 
$y_u = O(10^{-5})$ and $y_c = O(10^{-2})$
in the case with $|{R_{11}}^{ij}| = O(1)$, $|{R_{22}}^{ij}| = O(1)$,
and $|y^{(u)}_{ij}|=O(1)$ for their non-vanishing components.

\subsection{Reexamination of Yukawa coupling hierarchy}

Let us reexamine the hierarchy among up-type quark Yukawa couplings,
based on a general argument in Sect. 2.
We consider a simple case with three fundamental parameters 
$y$, $\varepsilon_1$, and $\varepsilon_2$,
and asuume that $y_{ij}^{(u)}$ are composed of three parts 
with much different magnitudes:
\begin{eqnarray}
y_{ij}^{(u)} = y^{u(1)}_{ij}(y)
+y^{u(\varepsilon_{1})}_{ij}(y, \varepsilon_1)
+y^{u(\varepsilon_{2})}_{ij}(y, \varepsilon_1, \varepsilon_2),
\label{yij(u)}
\end{eqnarray}
where $\varepsilon_1$ and $\varepsilon_2$ are tiny parameters, i.e.,
$y (=O(1)) \gg \varepsilon_1 (=O(\lambda^4)) \gg \varepsilon_2 (=O(\lambda^7))$.

In this case, the hierarchy can be generated 
without severe fine tuning,
in the following setting (a) -- (c).
\begin{itemize}
\item[(a)] The magnitude of non-vanishing components in $y^{u(1)}_{ij}(y)$ is at most $O(1)$,
the rank of $y^{u(1)}_{ij}(y)$ is one,
and the magnitude of non-zero eigenvalue is $O(1)$.
\item[(b)] The magnitude of non-vanishing components in 
$y^{u(\varepsilon_{1})}_{ij}(y, \varepsilon_1)$ is at most $O(\varepsilon_1)$.
The rank of $y^{u(1)}_{ij}(y)+y^{u(\varepsilon_{1})}_{ij}(y, \varepsilon_1)$ is two,
and the magnitude of non-zero eigenvalues are $O(1)$ and $O(\varepsilon_1)$.
\item[(c)] The magnitude of non-vanishing components in
$y^{u(\varepsilon_{2})}_{ij}(y, \varepsilon_1, \varepsilon_2)$ is at most $O(\varepsilon_2)$.
\end{itemize}

In fact, under reasonable assumptions such that
\begin{eqnarray}
&~& \left|\frac{\partial y^{u(\varepsilon_2))}}{\partial y}\right| = O(\varepsilon_2),~~
\left|\frac{\partial y^{u(\varepsilon_2))}}{\partial \varepsilon_1}\right| = O(\varepsilon_2),~~
\left|\frac{\partial y^{u(\varepsilon_2))}}{\partial \varepsilon_2}\right| = O(1),
\label{y-A1}\\
&~& \left|\frac{\partial y^{u(\varepsilon_1))}}{\partial y}\right| = O(\varepsilon_1),~~
\left|\frac{\partial y^{u(\varepsilon_1))}}{\partial \varepsilon_1}\right| = O(1),~~
\label{y-A2}
\end{eqnarray}
degrees of fine tuning for $y_u$ and $y_c$ are estimated as
\begin{eqnarray}
\hspace{-5mm}
&~& \varDelta_{y} y_u \equiv \left|\frac{\partial \ln y_{u}}{\partial \ln y}\right|= O(1),~~
\varDelta_{\varepsilon_1} y_u \equiv 
\left|\frac{\partial \ln y_{u}}{\partial \ln \varepsilon_1}\right| = O(\varepsilon_1),~~
\varDelta_{\varepsilon_2} y_u \equiv 
\left|\frac{\partial \ln y_{u}}{\partial \ln \varepsilon_2}\right| = O(1),
\label{yu-sen-fund}\\
\hspace{-5mm}
&~& \varDelta_{y} y_c \equiv \left|\frac{\partial \ln y_{c}}{\partial \ln y}\right|= O(1),~~
\varDelta_{\varepsilon_1} y_c \equiv 
\left|\frac{\partial \ln y_{c}}{\partial \ln \varepsilon_1}\right| = O(1),~~
\varDelta_{\varepsilon_2} y_c \equiv 
\left|\frac{\partial \ln y_{c}}{\partial \ln \varepsilon_2}\right| 
= O(\varepsilon_2/\varepsilon_1),
\label{yd-sen-fund}
\end{eqnarray}
and they suggest that the large hierarchy with
$y_t = O(1)$, $y_c = O(\varepsilon_1)$, 
and $y_u = O(\varepsilon_2)$
is naturally realized.

Finally, let us give an illustration with a matrix given by
\begin{eqnarray}
y_{ij}^{(u)} = y S_{ij} + y^{u(\varepsilon_1)}_{ij}(y, \varepsilon_1)
 + y^{u(\varepsilon_2)}_{ij}(y, \varepsilon_1, \varepsilon_2),
\label{yij(u)-ex}
\end{eqnarray}
where $S_{ij}$ is the $(i, j)$ component of the democratic matrix defined by
\begin{eqnarray}
S \equiv \frac{1}{3}
\left(
\begin{array}{ccc}
1 & 1 & 1 \\
1 & 1 & 1 \\
1 & 1 & 1
\end{array} 
\right).
\label{Democratic}
\end{eqnarray}
$S$ is easily diagonalized as
$USU^{\dagger} = {\rm diag}(0, 0, 1)$
with the unitary matrix:
\begin{eqnarray}
U = \frac{1}{\sqrt{3}}
\left(
\begin{array}{ccc}
\overline{\omega} & \omega & 1 \\
\omega & \overline{\omega} & 1 \\
1 & 1 & 1
\end{array}
\right),
\label{U}
\end{eqnarray}
where $\omega = e^{2\pi i/3}$ 
and $\overline{\omega} = \omega^2 = e^{4\pi i/3}$.
$y^{u(\varepsilon_{1})}_{ij}$ and $y^{u(\varepsilon_{2})}_{ij}$ are 
supposed to be given by polynomials of $\varepsilon_1$ and $\varepsilon_2$ such that
\begin{eqnarray}
y^{u(\varepsilon_1)}_{ij}(y, \varepsilon_1) = \sum_{n = 1} 
c_{ij,n}^{(\varepsilon_1)}(y)\varepsilon_1^{n},~~
y^{u(\varepsilon_2)}_{ij}(y, \varepsilon_1, \varepsilon_2) = \sum_{n = 1} 
c_{ij,n}^{(\varepsilon_2)}(y, \varepsilon_1)\varepsilon_2^{n}.
\label{yu-exp}
\end{eqnarray}
Then, $y^{(u)}_{ij}$ is diagonalized as
\begin{eqnarray}
V_{\rm L}^{(u)} \left(y S + y^{u(\varepsilon_1)}(y, \varepsilon_1)
+ y^{u(\varepsilon_2)}(y, \varepsilon_1, \varepsilon_2)\right)
{V_{\rm R}^{(u)}}^{\dagger}
= {\rm diag}\left(y_u, y_c, y_t\right),
\label{yu-diag-ex}
\end{eqnarray}
where $V_{\rm L}^{(u)}$, $V_{\rm R}^{(u)}$, $y_u$, $y_c$, and $y_t$ are
perturbatively given by 
\begin{eqnarray}
&~& (V_{\rm L}^{(u)})_{ij} = U_{ij} 
+ \sum_{n = 1}^{\infty} c_{ij,n}^{({\rm L},\varepsilon_1)}(y)
\varepsilon_1^{n}
+ \sum_{n = 1}^{\infty} c_{ij,n}^{({\rm L},\varepsilon_2)}(y, \varepsilon_1)
\varepsilon_2^{n},
\label{VL-ex}\\
&~& (V_{\rm R}^{(u)})_{ij} = U_{ij}
+ \sum_{n = 1}^{\infty} c_{ij,n}^{({\rm R},\varepsilon_1)}(y)
\varepsilon_1^{n}
+ \sum_{n = 1}^{\infty} c_{ij,n}^{({\rm R},\varepsilon_2)}(y, \varepsilon_1)
\varepsilon_2^{n},
\label{VR-ex}\\
&~& y_u = \sum_{n = 1}^{\infty} c^{(u, \varepsilon_2)}_n(y, \varepsilon_1) \varepsilon_2^{n},~~
y_c = \sum_{n = 1}^{\infty} c^{(c, \varepsilon_1)}_n(y) \varepsilon_1^{n}
+ \sum_{n = 1}^{\infty} c^{(c,\varepsilon_2)}_n(y, \varepsilon_1) \varepsilon_2^{n},
\label{yu-ex}\\
&~& y_t = y + \sum_{n = 1}^{\infty} c^{(t, \varepsilon_1)}_n(y) \varepsilon_1^{n}
+ \sum_{n = 1}^{\infty} c^{(t, \varepsilon_2)}_n(y, \varepsilon_1) \varepsilon_2^{n}.
\label{yt-ex}
\end{eqnarray}
Degrees of fine tuning for $y_u$ and $y_c$ are estimated as
\begin{eqnarray}
&~& \varDelta_{y} y_u = O(1),~~
\varDelta_{\varepsilon_1} y_u = O(\varepsilon_1),~~
\varDelta_{\varepsilon_2} y_u = O(1),
\label{yu-sen-fund-ex}\\
&~& \varDelta_{y} y_c = O(1),~~
\varDelta_{\varepsilon_1} y_c = O(1),~~
\varDelta_{\varepsilon_2} y_c = O(\varepsilon_2/\varepsilon_1),
\label{yc-sen-fund-ex}
\end{eqnarray}
using $|c^{(u, \varepsilon_2)}_1| = O(1)$,
$|\partial c^{(u, \varepsilon_2)}_1/\partial y| = O(1)$,
$|\partial c^{(u, \varepsilon_2)}_1/\partial \varepsilon_1| = O(1)$,
$|c^{(c, \varepsilon_1)}_1| = O(1)$,
$|c^{(c, \varepsilon_2)}_1| = O(1)$,
$|\partial c^{(c, \varepsilon_1)}_1/\partial y| = O(1)$,
$|\partial c^{(c, \varepsilon_2)}_1/\partial y| = O(1)$, and
$|\partial c^{(c, \varepsilon_2)}_1/\partial \varepsilon_1| = O(1)$.
From Eqs.~(\ref{yu-sen-fund-ex}) and (\ref{yc-sen-fund-ex}),
there seems no severe fine tuning to derive $y_u$ and $y_c$.
Note that no contributions of $y_{ij}^{u(1)}(y) = y S_{ij}$ in $y_u$ and $y_c$ stem from 
$(USU^{\dagger})_{11} = 0$ and $(USU^{\dagger})_{22} = 0$
relating to ${\rm S}_3$ symmetry.

\section{Conclusions and discussions}

We have reconsidered a generation of
up-type quark mass hierarchy in the SM,
and clarified how a mechanism works 
that the hierarchy is realized without severe fine tuning.
Based on the idea that
parameters in an effective QFT are not necessarily fundamental but effective ones,
and a tininess of physical parameters
can be naturally understood by using more fundamental parameters,
we have found that up-type quark mass hierarchy
can be naturally realized, if up-type Yukawa coupling matrix
consists of several parts with much different magnitudes,
the rank of a dominant part is one,
and the rank of a sum of dominant and semi-dominant ones is two.

The mechanism is available for the generation of both up-type and down-type quark
mass hierarchies in an extension of the SM with extra vector-like fermions.
We consider a case with $n$ families of quarks and $n-3$ families of mirror quarks.
It is assumed that $n \times n$ up-type Yukawa coupling matrix
consists of $y^{u(1)}(=O(1))$ and other tiny ones,
and $n \times n$ down-type Yukawa coupling matrix also
consists of $y^{d(1)}(=O(1))$ and other tiny ones.
If the rank of the dominant part $y^{u(1)}$ is $n-2$ 
and that of $y^{d(1)}$ is $n-3$, 
there can appear two up-type quarks and three down-type quarks
much below the weak scale.
The $n-3$ sets of up-type and down-type quarks form
vector-like heavy fermions in company with mirror ones.
Then an up-type quark with a Yukawa coupling of $O(1)$
remains as a chiral one, it acquires a mass of the weak scale
after the breakdown of electroweak symmetry,
and it is identified as a top quark.

It would be interesting to study a mechanism 
behind the flavor structure,
from the aspect of an exploration of a theory beyond the SM.
As a by-product, we might close in on an unknown part of QFTs
through generic features in the mechanism.

\section*{Acknowledgments}
This work was supported in part by scientific grants 
from the Ministry of Education, Culture,
Sports, Science and Technology under Grant No.~17K05413.

\end{document}